\documentclass[prl,twocolumn,showpacs,floatfix,superscriptaddress,longbibliography]{revtex4-1}
\usepackage{mathrsfs,braket}
\usepackage{amssymb, amsbsy, amsmath, latexsym, dsfont, array, layout, graphicx, mathrsfs, color, ulem, bm}
\usepackage[colorlinks,linkcolor=blue,anchorcolor=red,citecolor=blue]{hyperref}

\begin{document}

\title{Topological Sensing in the Dynamics of Quantum Walks with Defects}

\author{Xiaowei Tong}
\affiliation{School of Physics, Southeast University, Nanjing 211189, China}
\affiliation{Key Laboratory of Quantum Materials and Devices of Ministry of Education, Southeast University, Nanjing 211189, China}
\author{Xingze Qiu}
\affiliation{School of Physics Science and Engineering, Tongji University, Shanghai 200092, China}
\author{Xiang Zhan}
\affiliation{School of Physics, Southeast University, Nanjing 211189, China}
\affiliation{Key Laboratory of Quantum Materials and Devices of Ministry of Education, Southeast University, Nanjing 211189, China}
\author{Quan Lin}
\affiliation{School of Physics, Southeast University, Nanjing 211189, China}
\affiliation{Key Laboratory of Quantum Materials and Devices of Ministry of Education, Southeast University, Nanjing 211189, China}
\author{Kunkun Wang}\email{kunkunwang@126.com}
\affiliation{School of Physics and Optoelectronic Engineering, Anhui University, Hefei 230601, China}
\author{Franco Nori}\email{fnori@riken.jp}
\affiliation{RIKEN Quantum Computing Center, RIKEN, Wakoshi, Saitama, 351-0198, Japan}
\affiliation{Physics Department, The University of Michigan, Ann Arbor, Michigan 48109-1040, USA}
\author{Peng Xue}
\email{gnep.eux@gmail.com\\Present address: Beijing Computational Science Research Center, Beijing 100193, China.}
\affiliation{School of Physics, Southeast University, Nanjing 211189, China}


\begin{abstract}
Topological quantum sensing leverages unique topological features to suppress noise and improve the precision of parameter estimation, emerging as a promising tool in both fundamental research and practical application. In this Letter, we propose a sensing protocol that exploits the dynamics of topological quantum walks incorporating localized defects. Unlike conventional schemes that rely on topological protection to suppress disorder and defects, our protocol harnesses the evolution time as a resource to enable precise estimation of the defect parameter. By utilizing topologically nontrivial properties of the quantum walks, the sensing precision can approach the Heisenberg limit. We further demonstrate the performance and robustness of the protocol through Bayesian estimation. Our results show that this approach maintains high precision over a broad range of parameters and exhibits strong robustness against disorder, offering a practical pathway for topologically enhanced quantum metrology.

\end{abstract}

\maketitle
{\it Introduction.---}Achieving high-precision sensing is essential for advancing both fundamental scientific research and practical applications. Quantum-enhanced sensing has shown the potential to beat the standard quantum limit and approach the Heisenberg limit~\cite{1giovannetti2004quantum,2giovannetti2006quantum,3degen2017quantum,4wang2018entanglement,5huang2024entanglement,6shen2025entanglement}, as demonstrated by implementations using Greenberger-Horne-Zeilinger (GHZ) states~\cite{7che2019multiqubit,8li2020multipartite,9cao2024multi,10yin2024heisenberg, 11yin2025fast}, N00N states~\cite{12lee2002quantum,13joo2011quantum,LY23,14liu2024quantum}, Fock states~\cite{15wang2019heisenberg,16tsai2023fisher,17deng2024quantum,18rahman2025genuine}, and squeezed states~\cite{19hu1996squeezed,20zuo2020quantum,21cardoso2021superposition,22xin2021rapid,xu2022metrological,23malitesta2023distributed,24qin2024exponentially,25zhang2024squeezing,qin2024enhanced,zhang2024quantum}. Despite these proven advantages, the practical implementation of quantum-enhanced sensing is often hindered by experimental challenges. These include the difficulties in preparing entangled states and their susceptibility to environmental noise, both of which can degrade the enhanced sensing precision.

A promising solution is to exploit the topological properties of quantum systems to achieve a robust sensing protocol. Topological invariants, such as Chern numbers~\cite{26wang2012simplified,27fukui2005chern} or winding numbers~\cite{28pezze2017multipartite,29zhang2018characterization,30yin2018geometrical,31wang2021detecting}, ensure that the associated band structures or edge states are robust against impurities, minor defects, and fabrication imperfections. This inherent protection allows the system to retain key properties, such as propagation direction, energy, and conductivity, demonstrating inherent structural robustness~\cite{32barik2016two,33barik2018topological,34wang2020terahertz}. As a result, topologically nontrivial systems provide a robust foundation for sensing protocols capable of operating reliably under noise and disorder. Such topological sensing protocols have been investigated, both theoretically and experimentally, across diverse physical platforms, including photonic crystals~\cite{35bandres2016topological,36weimann2017topologically,37arledge2021topological,38cheng2022topological,39khanikaev2024topological,bao2024exponential}, phononic systems~\cite{40wang2020robust,41wu2022topological,42zhao2024electrostatically}, cold atoms~\cite{43goldman2016topological,44sarkar2022free,45euler2023detecting}, etc.

\begin{figure}[t]
\includegraphics[width=0.45\textwidth]{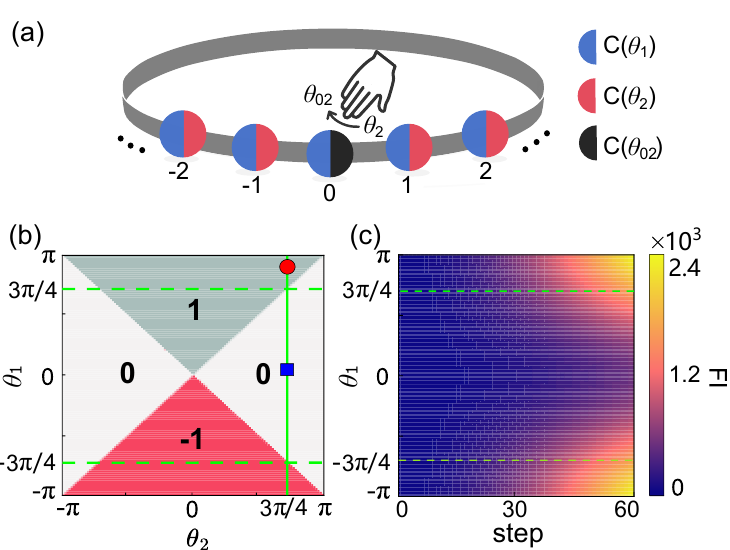}
\caption{(a) Schematic illustration of a split-step quantum walk (QW) with a localized defect. Driven by two shift and coin operators with parameters $\theta_1$ and $\theta_2$, a defect is introduced by changing the coin parameter $\theta_2$, at position 0, to $\theta_{02}$. (b) Topological phase diagram, characterized by a winding number as a function of coin parameters $(\theta_1, \theta_2)$. The red circle and the blue square denote two specific choices of $(\theta_1, \theta_2)$, corresponding to the topologically nontrivial and trivial phases, respectively. (c) Fisher information (FI) versus the evolution time and $\theta_1$, and $\theta_{2}$ is fixed to $0.75\pi$. The green dashed lines mark the boundary between different phases.
}
\label{fig1}
\end{figure}

However, these protocols rely on either large-scale systems or those with multiple degrees of freedom to approach the Heisenberg limit~\cite{44sarkar2022free,45euler2023detecting}, and their sensitivity tends to peak sharply near topological phase transition points~\cite{46liu2016metrology,47wu2021criticality,48liu2021experimental,49bao2021fundamental,50yang2023spectral,51qin2024quantum,52parto2025enhanced,beaulieu2025criticality}. Thus, their applicability is generally confined to narrow regions around these critical points. Moreover, it relies on specific structural designs to introduce topological protection, enhancing robustness and suppressing measurement accuracy degradation from defects.


In this letter, we propose a robust and wide-range high-precision sensing protocol based on a topological quantum walk (QW) model. Notably, our potocol achieves Heisenberg-limited precision by exploiting the evolution time, quantified by the number of steps in the QW, as a quantum resource. Distinct from conventional methods that suppress defects through topological protection, our approach incorporates the defect into a QW. Specifically, the defect is introduced by modifying the coin parameter at a specific position, which is regarded as the unknown parameter to be estimated. Through a detailed analysis of Fisher information (FI)~\cite{53ma2011quantum} and Bayesian estimation within the dynamics of a split-step QW, we quantitatively characterize the parameter estimation capabilities of our protocol. The results reveal \textit{Heisenberg-limited sensitivity across a broad parameter range and pronounced robustness against disorder} in the topologically nontrivial regime. Our protocol circumvents the need for large-scale resources or complex state preparation, thereby making it both practically accessible and resilient in real-world applications.

{\it Model.---}We consider a split-step QW~\cite{qiang2016efficient,53cardano2016statistical,54ramasesh2017direct,55xiao2017observation,su2019experimental,yan2019strongly,56shukla2024quantum,57meng2024generation,58feis2025space,59huo2025experimental} on a 1D lattice, which allows the system to manifest a variety of topological phases.
The state of the walker is represented as \(|x,c\rangle=|x\rangle\otimes|c\rangle \), where \(|x\rangle\) denotes the position state with \(x\in\mathbb{Z}\), and \(|c\rangle\) denotes the coin state with \(c\in\{\uparrow, \downarrow\}\).
Each step of the evolution is governed by a unitary operator \(U=T_{\downarrow}\tilde{R}_2 T_{\uparrow}\tilde{R}_1\), which consists of two shift operators, \({T}_\uparrow\) and \({T}_\downarrow\), and coin operators, $\tilde{R_1}$ and $\tilde{R_2}$.

The shift operators are defined as \begin{equation}
\begin{aligned}
T_{\uparrow} &= \sum_{x} \ket{x+1}\bra{x} \otimes \ket{\uparrow}\bra{\uparrow} + \ket{x}\bra{x} \otimes \ket{\downarrow}\bra{\downarrow}, \\
T_{\downarrow} &= \sum_{x} \ket{x-1}\bra{x} \otimes \ket{\downarrow}\bra{\downarrow} + \ket{x}\bra{x} \otimes \ket{\uparrow}\bra{\uparrow}.
\end{aligned}
\end{equation} The coin operators are given by \(\tilde{R_i}=\sum_x|x\rangle\langle x|\otimes {R} (\theta_i)\) for \(i\in\{1, 2\}\), where the coin flipping operator ${R} (\theta_i)=e^{-i\theta_i\sigma_y/2}$, with $\sigma_y$ being the Pauli Y matrix. As shown in Fig.~\ref{fig1}(a), we implement the model under periodic boundary conditions with uniform coin operators.
The parameter to be estimated enters the system via the coupling between the QW and the external object at a specific position, e.g., $x=0$. This coupling modifies the coin operator, introducing a local defect in the second flipping operator $R(\theta_{02})$, while the coin operators at all other sites remain unchanged.


The topological properties of the QW can be characterized by a winding number, as shown in the phase diagram for the QW in Fig.~\ref{fig1}(b). The winding number~\cite{30yin2018geometrical, 61Supplemental Material} is defined as the number of times the Hamiltonian trajectory encircles the origin in parameter space as momentum spans the Brillouin zone,
\begin{equation}\nu=-\frac{1}{2\pi}\int_{-\pi}^{\pi} \mathbb{A}\cdot\left(\boldsymbol{n}(k)\times\frac{\mathrm{d}\boldsymbol{n}(k)}{\mathrm{d}k}\right).\end{equation}
Here $\boldsymbol{n}(k)$ denotes the unit vector of the effective Hamiltonian, and $\mathbb{A}$ is a fixed unit vector chosen perpendicular to the Bloch vector. 

\begin{figure}
  \centering
  \includegraphics[width=0.5\textwidth]{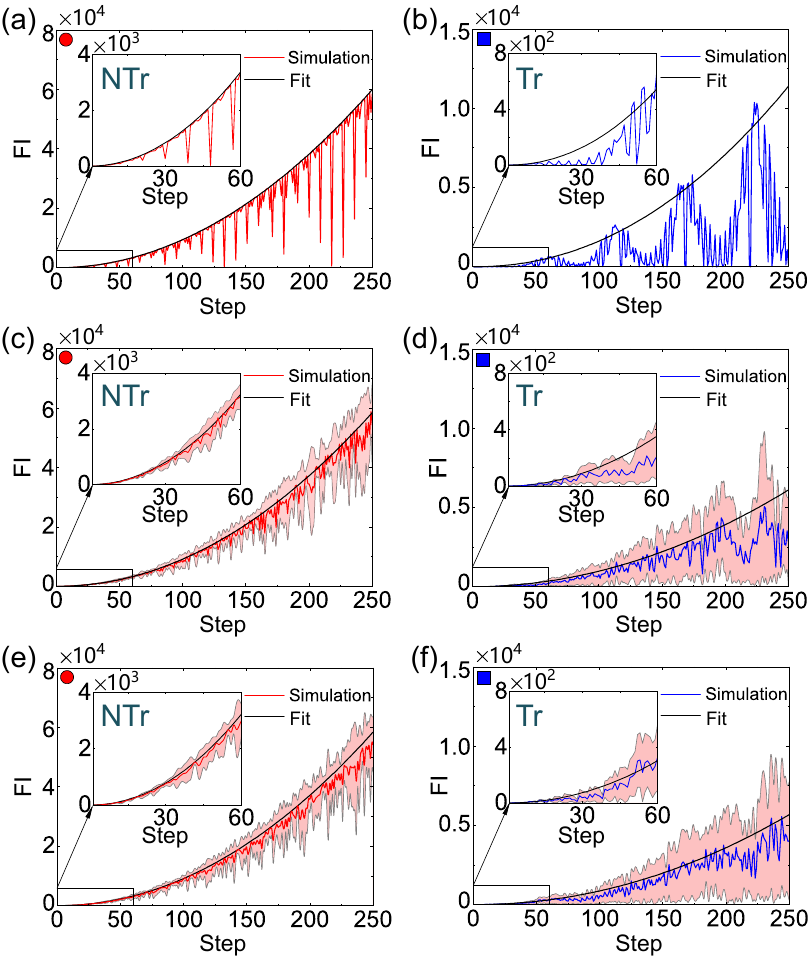}
  \caption{The colored curves depict the dynamical evolution of the FI over steps with fixed parameters $\theta_2 = 0.75\pi$ and $\theta_{02} = -0.55\pi$. In panel (a), $\theta_1 = 0.90\pi$ is chosen within the topologically nontrivial (NTr) regime, while in panel (b), $\theta_1 = 0.05\pi$ corresponds to the topologically trivial (Tr) regime. The black solid curve represents the Heisenberg limit scaling. (c) and (d) show the variation of the FI under static disorder. The coin parameters are randomly chosen in the interval $[\theta_i-\pi/20, \theta_i+\pi/20]$ at each position. (e) and (f) show the variation of the FI under dynamic disorder. The coin parameters are randomly chosen in the interval $[\theta_i-\pi/20, \theta_i+\pi/20]$ at each step. The colored curves in (c-f) correspond to the data obtained by averaging over 10 different disorder configurations, with the red shaded area representing the corresponding standard deviation.}
  \label{fig2}
\end{figure}

Under periodic boundary conditions, the QW model preserves translational symmetry. When a defect is introduced (i.e., when \(\theta_{02}\neq\theta_2\)), an effective boundary is artificially created. This defect breaks the translational symmetry and locally modifies the scattering or transition rules near the defect position. The larger the difference between the bulk parameter $\theta_2$ and the defect parameter $\theta_{02}$, the more pronounced the effective boundary becomes, leading to two localized states around the defect, with their amplitudes decaying exponentially into the bulk \cite{61Supplemental Material}. When the system is in a topologically nontrivial phase, the localized states inherit topologically protected features from the bulk and exhibit robustness against disorder. This indicates that the sensing protocol based on two defect-induced localized states benefits from topological protection, which is a key feature of our topological quantum sensing scheme.




{\it Sensing precision.---}We aim to estimate the parameters of the defect with robustness and high precision. The precision for estimating a single unknown parameter is typically quantified by the statistical standard deviation, which is lower-bounded by the well-known Cram\'er-Rao bound~\cite{62braunstein1994statistical,63hradil1996quantum,64pezze2007phase}, i.e., $\Delta\theta_{02}^2\geq1/\text{FI}$. Here FI represents the classical FI~\cite{65barndorff2000fisher,66paris2009quantum,67wang2014quantum,68liu2017quantum,69liu2020quantum,lee2023steering,70yu2024toward,rinaldi2024parameter,71gorecki2025mutual}, which can be expressed as a function of the number of steps \textit{t} of the QW in our protocol\begin{equation}
\text{FI}(t)=[\partial P_0(t)/\partial \theta_{02}]^2/\{P_0(t)[1-P_0(t)]\},
\end{equation}  where $P_0(t)$ denotes the probability of finding the walker at the defect site $x = 0$ after $t$ steps of the quantum walk characterized by the defect parameter $ \theta_{02}$. The initial state is set to $|-1,\downarrow\rangle$. In general, the FI increases with the number of steps $t$, following a scaling $\mathrm{FI}\sim t^{b}$. In the classical shot-noise limit, the maximum achievable scaling exponent is $b=1$. In quantum-enhanced measurements, the ultimate limit is $b=2$, known as the Heisenberg limit. In the following analysis, we characterize the properties of our quantum-enhanced sensing protocol by comparing both the amplitude and the scaling exponent of the FI under topologically nontrivial and trivial conditions.

We fix the coin parameters along the green solid line shown in Fig.~\ref{fig1}(b), where one of the coin parameters is set to $\theta_2=0.75\pi$ and the defect parameter is fixed to $\theta_{02}=-0.99\pi$. As shown in Fig.~\ref{fig1}(c), by varying $\theta_1$, we obtain the dynamical FI across three distinct topological phases, which are separated by the horizontal dashed green lines. The upper and lower regions correspond to two topologically nontrivial invariants $\pm1$, while the central region represents the topologically trivial case. The FI increases with the number of steps and is significantly greater in the topologically nontrivial regions compared to the topologically trivial region. This suggests that the topological sensing property characterized by the FI can also serve as an effective indicator of topological phase transitions. Moreover, the FI exhibits identical behavior as a function of the number of steps in both topologically nontrivial regions with winding numbers $+1$ and $-1$. This symmetry is clearly illustrated in Fig.~\ref{fig1}(c), confirming the numerical equivalence of their values. In the following, we focus on one of the topologically nontrivial cases (with winding numbers $+1$) to illustrate its key properties and related comparisons.
\begin{figure}[hbtp]
\includegraphics[width=0.5\textwidth]{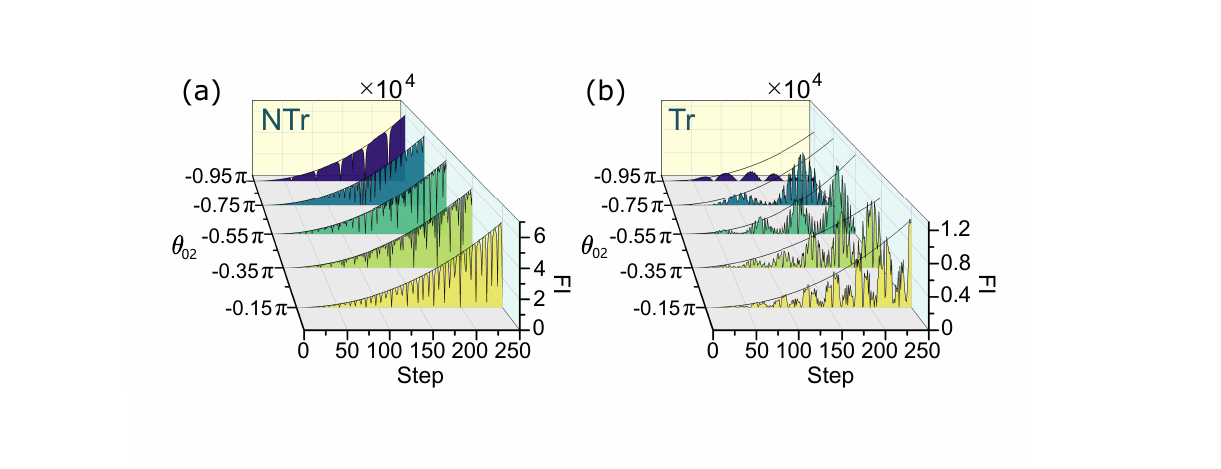}
\caption{FI under different values of the defect parameter $\theta_{02}$. (a) Topologically nontrivial case. (b) Topologically trivial case. The black solid lines represent the fitting curve corresponding to quadratic scaling proportional to $t^2$ (Heisenberg limit). The parameters are fixed to $\theta_2=0.75\pi$, with $\theta_1=0.90\pi$ in (a) and $\theta_1=0.05\pi$ in (b).}
\label{fig3}
\end{figure}

We now consider two sets of parameters, lying in the topologically nontrivial and trivial regions, indicated by the red circle and the blue square in Fig.~\ref{fig1}(b), respectively. By examining the FI and its scaling behavior with respect to the evolution time, we perform a detailed analysis and comparison of the sensing performance in these two cases. As shown in Fig.~\ref{fig2}(a), for the topologically nontrivial case with $\theta_1=0.9\pi$ and $\theta_2=0.75\pi$, the FI grows quadratically with the number of steps $t$. We fit the scaling of the FI with respect to the number of steps using the relation $\mathrm{FI} \sim t^b$. The resulting exponent, $b \approx 2$, indicates that the protocol achieves Heisenberg-limited sensitivity.

In the topologically trivial case with $\theta_1=0.9\pi$ and $\theta_2=0.75\pi$, as shown in Fig.~\ref{fig2}(b), the FI is generally much smaller than that in the topologically nontrivial case. Moreover, the FI in the topologically trivial case displays a distinguishable oscillatory behavior. To qualitatively analyze this behavior, we extract the peak values of the FI and fit them using the same scaling relation $\mathrm{FI}\sim t^{b}$. Although the fit again yields $b\approx2$, only a few points near the oscillation peaks approach the Heisenberg limit, while the majority remain significantly below it.

To verify the robustness of the sensing protocol for these two cases, we introduce two different types of disorder to the dynamics of the QW. Static disorder is introduced by randomly choosing the coin parameters from the interval $[ \theta_i-\pi/20, \theta_i+\pi/20 ]$ at each position as shown in Figs.~\ref{fig2}(c-d). This type of disorder is spatially dependent but remains constant from step to step. Dynamic disorder is implemented by randomly varying the coin parameters at each step and uniformly across all positions, within the same interval as shown in Figs.~\ref{fig2}(e-f). The results show that under both types of disorder, the disorder-averaged FI in the topologically nontrivial region remain consistent with the original distribution along the Heisenberg limit, which demonstrates the robustness. In contrast, in the topologically trivial region, the FI loses its original scaling behavior and exhibits larger disorder-induced fluctuation compared to the topologically nontrivial case. These results clearly demonstrate the robustness of topologically sensing against disorder.

\begin{figure*}[ht!]
\includegraphics[width=1\textwidth]{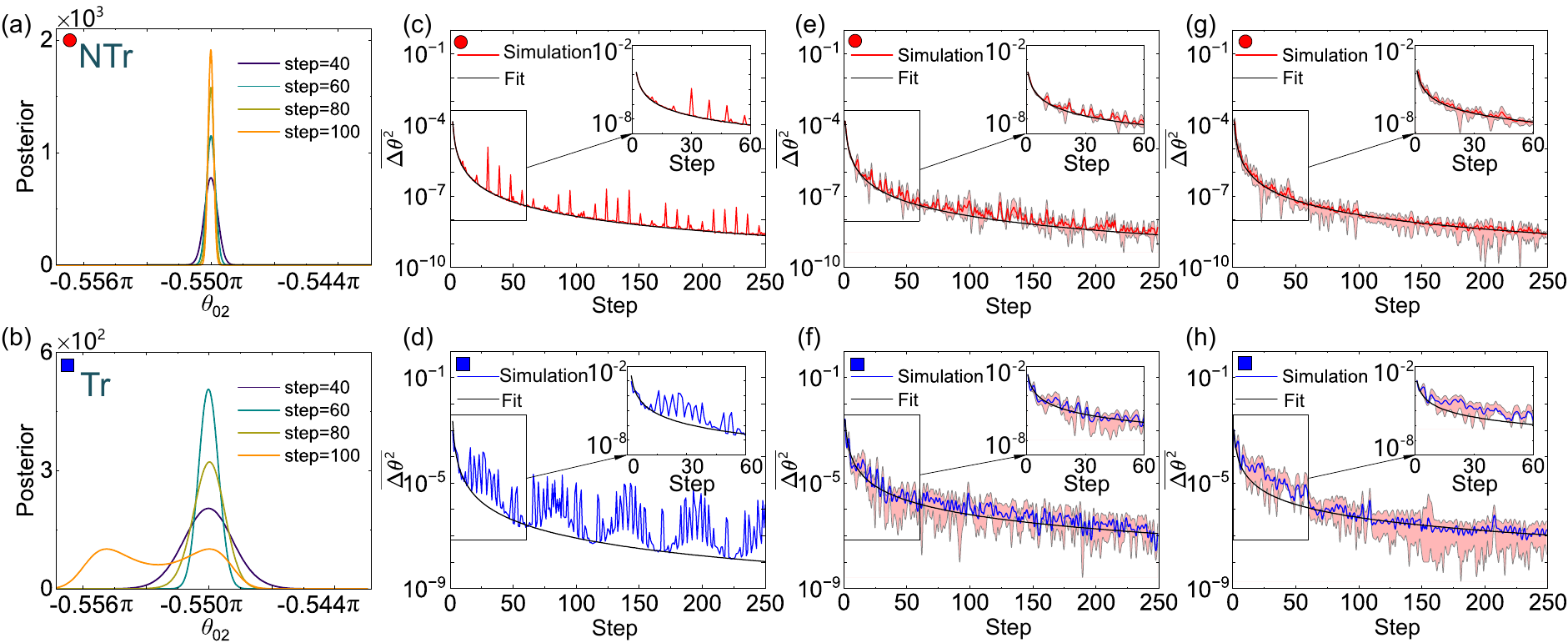}
\caption{Numerical results of Bayesian parameter estimation. The posterior distributions for sensing $\theta_{02}=-0.55\pi$ with uniform priors $\theta_{02}\in[-0.556\pi, -0.544\pi]$ in topologically nontrivial (a) and trivial (b) cases, respectively. (c) and (d) show the mean of the squared relative error $\Delta \theta_{02}^2$ as a function of steps in two cases. The black solid line represents the Heisenberg limit. The red curve shows the result for the topologically nontrivial case with fixed parameters $\theta_1=0.90\pi$ and $\theta_2=0.75\pi$, while the blue curve corresponds to the topologically trivial case with $\theta_1=0.05\pi$ and $\theta_2=0.75\pi$. (e) and (f) show the impact of static disorder to $\Delta \theta_{02}^2$. (g) and (h) show the impact of dynamic disorder. The colored curves in (e-h) correspond to the data obtained by averaging over 10 different disorder configurations, with the red shaded area representing the corresponding standard deviation.
}
\label{fig4}
\end{figure*}

As shown in Fig.~\ref{fig3}, we fix the parameters $\theta_{1}$ and $\theta_{2}$, and investigate whether the protocol can maintain high sensing precision over a wide range of defect parameters to be estimated. We calculate the FI for a set of representative values of $\theta_{02}$ to evaluate the sensing precision. In the topologically nontrivial case, we find that high-precision sensing is preserved over a broad range of $\theta_{02}$, with the FI consistently exhibiting the Heisenberg scaling, i.e., $b\approx2$. Defining the defect strength as the difference between the bulk parameter $\theta_2$ and the defect parameter $\theta_{02}$, we observe that, for the topologically nontrivial case, the FI increasingly concentrates around the Heisenberg limit as the defect strength grows. By contrast, in the topologically trivial case, the region exhibiting Heisenberg scaling decreases as the defect strength increases. While our main comparison considers parameters deep in the topologically nontrivial and trvial phases, the same results are also obtained near the topological phase transition point, as demonstrated in~\cite{61Supplemental Material}.

{\it Parameter estimation.---}Since the FI is difficult to measure directly in experiments, we employ Bayesian estimation~\cite{72dinani2019bayesian} with uniform prior to evaluate Heisenberg-limited precision. In this framework, the Cram\'er-Rao bound is interpreted in terms of estimation error. The Bayesian estimation is based on Bayes rule, $P(d|D)=P(D|d)P(d)/P(D)$, where $P(d)$ represents the prior distribution of the unknown parameter. In our protocol, we assume a uniform prior $P(d)$. Note that for non-uniform priors, the Ziv-Zakai bound~\cite{73bell1997extended} is preferable, as it outperforms the Cram\'er-Rao bound. The marginal distribution of the data, $P(D)$, serves as a normalization factor. Since both $P(d)$ and $P(D)$ can be treated as constants, the posterior distribution $P(d|D)$ is proportional to the likelihood function $P(D|d)$.


We repeat the dynamics of the QW for $M$ times while measuring the appearance of the walker at the defect position. The probability of detecting the walker at the defect position $m$ times follows a binomial distribution. The likelihood function $P(D|d)$ of the distribution, where $d$ represents the coin parameter at the defect position to be estimated, can be expressed accordingly \cite{74pezze2018quantum}\begin{equation}\begin{aligned}
P(D|d)=\binom{M}{m}P_1^m(t)[1-P_1(t)]^{M-m}.
\end{aligned}\end{equation}
Note that the posterior distribution satisfies the relation $P(d|D)\propto P(D|d)$. In our sensing protocol, the parameter of interest is $d=\theta_{02}$ and $D=\{\theta_1,\theta_2\}$. Therefore, the posterior distribution represents the probability distribution of $\theta_{02}$ conditioned on the known values of $\theta_{1}$ and $\theta_{2}$. When the number of measurements $M$ is sufficiently large, the posterior converges to a Gaussian distribution centered on the true value of the unknown parameter $\theta_{02}$. An excellent candidate for implementing Bayesian parameter estimation in the dynamics of the QW is the time-multiplexed scheme~\cite{75chalabi2019synthetic,76lin2023manipulating}, which is further discussed in~\cite{61Supplemental Material}.

Figure~\ref{fig4} shows the results for estimating the defect parameter $\theta_{02}$. As shown in Figs.~\ref{fig4}(a-b), we analyze the posterior distributions in both topologically nontrivial and trivial regions. In the topologically nontrivial region, the Gaussian wave packet converging around the true value becomes progressively narrower with increasing steps, indicating enhanced sensing precision. In contrast, the width of the Gaussian packet is significantly broader and varies irregularly with the number of steps in the topologically trivial region. These results highlight the superior sensing performance of the topological sensing protocol in the topologically nontrivial case.
We quantitatively analyze the impact of the linewidth of the Gaussian waveform on the parameter estimation performance. The mean squared relative error is used as the evaluation metric $\Delta \theta_{02}^2=(\sigma_{\theta_{02}}^2+|\langle \theta_{02}\rangle-\theta_{02}|^2)/{|\theta_{02}|^2}$. Here $\sigma_{\theta_{02}}$ and $\langle \theta_{02}\rangle$ are the variance and the average of $\theta_{02}$ with respect to the posterior distribution $P(\theta_{02}|\theta_1,\theta_2)$. The estimation accuracy achieved by this method satisfies the Cram\'er-Rao bound. Accordingly, the mean squared relative error corresponds to the inverse of the FI and exhibits a scaling relation $\Delta\theta_{02}^2 \sim t^{-2}$. In Figs.~\ref{fig4}(c-d), we present the variation of the mean squared relative errors with the number of steps for both two cases. By comparing with Figs.~\ref{fig2}(a-b), the results show great agreement with the behavior of the FI discussed earlier. Specifically, the error in the topologically nontrivial case is approximately one order of magnitude smaller than that in the topologically trivial case. In terms of the scaling exponent, the error in the topologically nontrivial region closely follows the Heisenberg limit, indicating a higher measurement precision.

In Figs.~\ref{fig4}(e-h), we introduce the static and dynamic disorder discussed previously to evaluate the robustness of the estimating protocol. The results show that, the error in the topologically nontrivial region remain consistent with the Heisenberg scaling, maintaining high sensing precision despite the presence of disorder. In contrast, in the topologically trivial region, the mean squared relative error deviates significantly from its original trend. Specifically, the mean squared relative error becomes significantly larger and more unstable due to the disorder, further highlighting the robustness advantage of the topological sensing protocol in the topologically nontrivial region.

{\it Conclusion.---}In this letter, we present an approach to topological quantum sensing by utilizing the evolution time as a key resource in the dynamics of the QW. Unlike conventional schemes that seek the topological features to suppress a defect, our protocol explicitly incorporates the defect into the QW model and treats it as the parameter to be estimated. Through comprehensive analyses using both FI and Bayesian estimation, we show that the sensing precision in the topologically nontrivial region achieves the Heisenberg limit. Moreover, compared to most topological sensing schemes that rely on quantum critical points and are only precise near such points, our protocol maintains Heisenberg-limited precision across a broad parameter range, enabling the construction of a highly sensitive and robust quantum sensor. The sensor also maintains strong resilience against the presence of both static and dynamic disorder, highlighting its practical robustness for real-world applications.

Our findings highlight that topological protection combined with defect engineering not only enhances robustness but also enables high-precision sensing across a wide parameter range. This paradigm redefines defects as beneficial elements rather than detrimental features, providing a practical and scalable route toward robust quantum metrology. As quantum sensing is widely recognized as a cornerstone for both fundamental discoveries and emerging quantum technologies, our results provide a meaningful step toward realizing highly sensitive, resilient, and experimentally feasible quantum sensors.

\begin{acknowledgments}
This work is supported by the National Key R\&D Program of China (Grant No. 2023YFA1406701) and the National Natural Science Foundation of China (Grants No. 12025401, No. 92265209, No. 12474352, No. 12088101). KKW acknowledges support from Beijing National Laboratory for Condensed Matter Physics (No. 2024BNLCMPKF010). FN is supported in part by: the Japan Science and Technology Agency (JST) [via the CREST Quantum Frontiers program Grant No. JPMJCR24I2, the Quantum Leap Flagship Program (Q-LEAP), and the Moonshot R\&D Grant Number JPMJMS2061].
\end{acknowledgments}

{\it Data availablity.---}The data that support the findings of this letter are not
publicly available. The data are available from the authors upon reasonable request.


\clearpage
\begin{widetext}
\appendix

\renewcommand{\thesection}{\Alph{section}}
\renewcommand{\thefigure}{S\arabic{figure}}
\renewcommand{\thetable}{S\Roman{table}}
\setcounter{figure}{0}
\renewcommand{\theequation}{S\arabic{equation}}
\setcounter{equation}{0}

\section{SUPPLEMENTAL MATERIALS FOR ``TOPOLOGICAL SENSING IN THE DYNAMICS OF QUANTUM WALKS WITH DEFECTS''}

In this Supplemental Materials section, we derive the associated topological invariant and analyze the influence of defects within a quantum walk (QW). Moreover, we propose a feasible experimental scheme based on the previously discussed model.

\subsection{Winding number}

As mentioned in the main text, in a one-dimensional discrete-time QW, the operator of $U=T_{\downarrow}\tilde{R}_2 T_{\uparrow}\tilde{R}_1$ describes the evolution of the QW over each step. Since all the operators can be expanded in the Pauli basis, the time-evolution operator can be rewritten in momentum space
\begin{equation}
\begin{aligned}
    U(k) &=d_0(k)\sigma_0+id_x\sigma_x+id_y(k)\sigma_y+id_z(k)\sigma_z,\\
    d_0(k) &=\cos\frac{\theta_2}{2}\cos\frac{\theta_1}{2}\cos k -\sin\frac{\theta_2}{2}\sin\frac{\theta_1}{2},\\
    d_x(k) &=\cos\frac{\theta_2}{2}\sin\frac{\theta_1}{2}\sin k,\\
    d_y(k) &=\cos\frac{\theta_2}{2}\sin\frac{\theta_1}{2}\cos k+\sin\frac{\theta_2}{2}\cos\frac{\theta_1}{2},\\
    d_z(k) &=-\cos\frac{\theta_2}{2}\cos\frac{\theta_1}{2}\sin k.
\end{aligned}
\end{equation}

\begin{figure*}[htbp]
    \centering
    \includegraphics[width=0.8\textwidth]{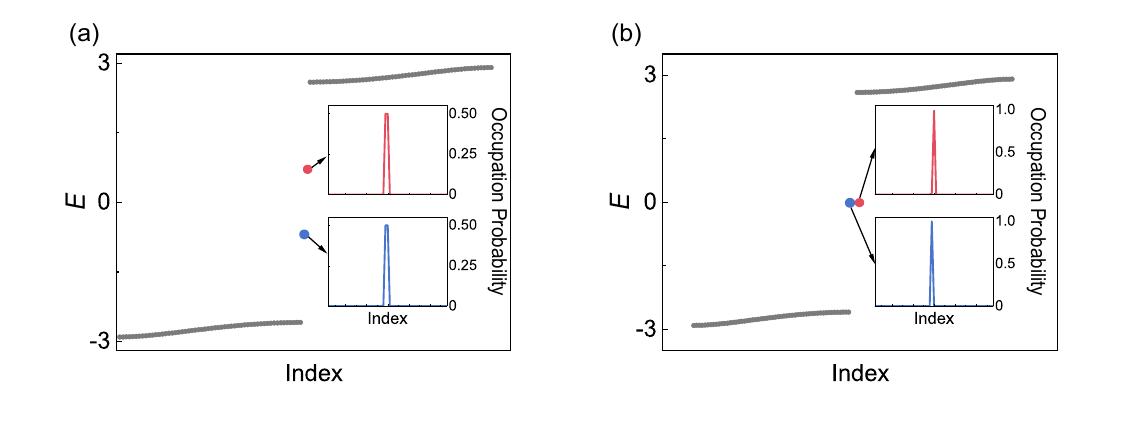}
    \caption{Energy bands and occupation probabilities (in the insets) of two defect-induced localized states with the fixed parameters $\theta_1 = 0.9\pi$, $\theta_2 = 0.75\pi$, and $\theta_{02} = -0.55\pi$ in (a), and $\theta_{02}=-\pi$ in (b). When $\theta_{02}$ approaches $-\pi$, the defect acts as a domain wall to interrupt the propagation, and the two localized states can be regarded as edge states.}
    \label{figS1}
\end{figure*}
\begin{figure*}[htbp]
    \centering
    \includegraphics[width=0.95\textwidth]{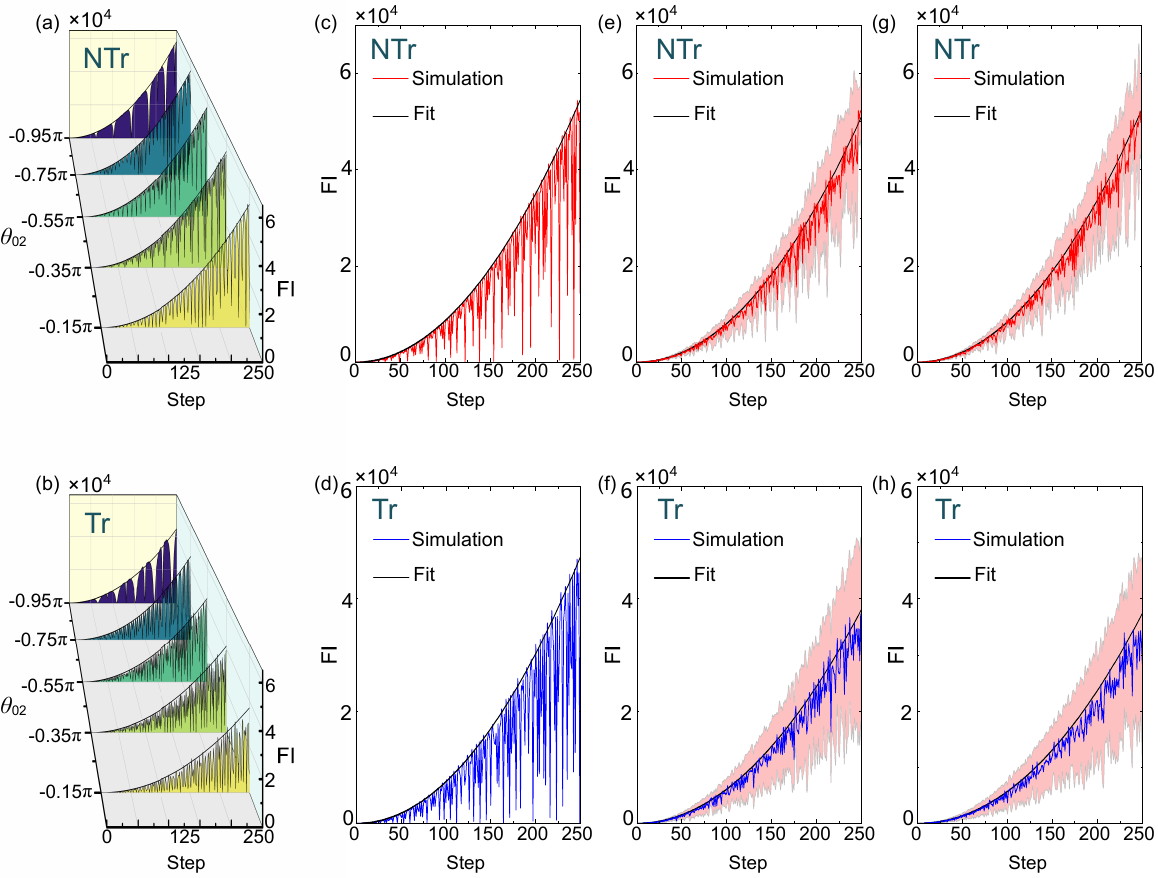}
    \caption{Fisher information (FI) for different values of the defect parameter $\theta_{02}$ near the topological phase transition point ($\theta_1=0.75\pi$, $\theta_2=0.75\pi$). In panel (a), $\theta_1 = 0.80\pi$ is selected within the topologically nontrivial (NTr) regime, while in panel (b), $\theta_1 = 0.70\pi$ lies in the topologically trivial (Tr) regime. In both cases, the parameter $\theta_2$ is fixed at $0.75\pi$. The black solid lines represent the fitting curves. In panels (c-h), the defect parameter is set to $\theta_{02}=-0.55\pi$. (c) Topologically nontrivial (NTr) case. (d) Topologically trivial (Tr) case. (e) and (f) show the variation of the FI under static disorder. The coin parameters are randomly chosen in the interval $[\theta_i-\pi/20, \theta_i+\pi/20]$ at each position. (g) and (h) show the variation of the FI under dynamic disorder. The coin parameters are randomly chosen in the interval $[\theta_i-\pi/20, \theta_i+\pi/20]$ at each step. The colored curves in (e-h) correspond to the data obtained by averaging over 10 different disorder configurations, with the red shaded area representing the corresponding standard deviation.}
    \label{figS2}
\end{figure*}
\noindent Here the $\boldsymbol{\sigma}=(\sigma_{x},\sigma_{y},\sigma_{z})$ ($\sigma_{x,y,z}$ are the Pauli matrices). The evolution $U$ is driven by a time-independent effective Hamiltonian $H$ with the relation $U=e^{-iH}$. In the quasi-momentum space, the effective Hamiltonian takes the form \begin{equation}
    H=\int_{-\pi}^{\pi}dk\left[E(k)\mathbf{n}(k)\cdot\boldsymbol{\sigma}\right]\otimes|k\rangle\langle k|.
\end{equation} The unit vector component of $H$ can be expressed as \begin{equation}\mathbf{n}(k)=\frac{1}{\sin[E(k)]}
\begin{pmatrix}
\cos\frac{\theta_2}{2}\sin\frac{\theta_1}{2}\sin k \\
\cos\frac{\theta_2}{2}\sin\frac{\theta_1}{2}\cos k+\sin\frac{\theta_2}{2}\cos\frac{\theta_1}{2}\\
-\cos\frac{\theta_2}{2}\cos\frac{\theta_1}{2}\sin k
\end{pmatrix}.\end{equation}
It can be straightfoward to check that $\mathbb{A}=\left\{\cos\frac{\theta_1}{2},0,\sin\frac{\theta_1}{2}\right\}$ is perpendicular to $\boldsymbol{n}(k)$. As the rotation component along the Hamiltonian direction is extracted by projecting onto $\mathbb{A}$,
the winding number is given by \begin{equation}\nu=-\frac{1}{2\pi}\int_{-\pi}^{\pi} \mathbb{A}\cdot\left(\boldsymbol{n}(k)\times\frac{\mathrm{d}\boldsymbol{n}(k)}{\mathrm{d}k}\right).\end{equation}

\subsection{properties of defect-induced localized states}

A defect in the quantum walk (QW) within the topologically nontrivial region under periodic boundary conditions effectively induces an artificial boundary. In Fig.~\ref{figS1}(a), we fix the defect parameter to $\theta_{02} = -0.55\pi$, resulting in the emergence of two localized states, which are exponentially localized at the position of the defect and possess eigenvalues of equal magnitude but opposite signs. The larger the difference between the bulk parameter $\theta_2$ and the defect parameter $\theta_{02}$, the more pronounced the effective boundary becomes. When the parameter of the defect tends to $\theta_{02}=-\pi$, the effect induced by this boundary can be regarded as nearly breaking the periodic boundary condition. As shown in Fig.~\ref{figS1}(b), the defect induces an effective domain wall, resulting in the localization of two edge states.

\begin{figure*}
    \centering
    \includegraphics[width=1\textwidth]{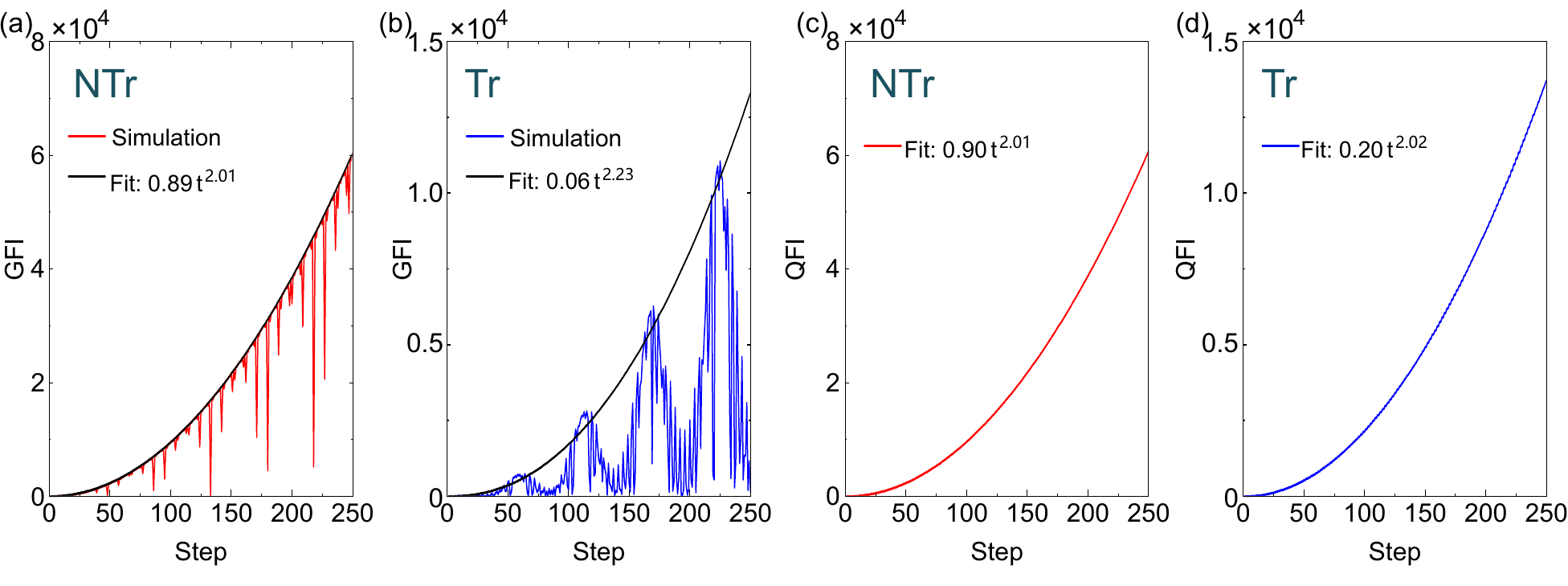}
    \caption{Fisher information for different measurement schemes. Panels (a) and (b) show the global Fisher information (GFI), while panels (c) and (d) show the quantum Fisher information (QFI), for the topologically nontrivial (NTr) and trivial (Tr) cases, respectively. The parameters are fixed to $\theta_2=0.75\pi$ and $\theta_{02}=-0.55\pi$, with $\theta_1=0.90\pi$ in (a) and (c), and $\theta_2=0.05\pi$ in (b) and (d).}
    \label{figS3}
\end{figure*}

\subsection{performance near the critical point}

To further compare the topologically nontrivial and trivial phases, we select two sets of parameters near the topological phase transition point ($\theta_1=0.75\pi$, $\theta_2=0.75\pi$). As shown in Fig.~\ref{figS2}, the parameters are chosen as $\theta_1=0.80\pi$ and $\theta_1=0.70\pi$ for the topologically nontrivial and trivial phases, respectively, with $\theta_2$ fixed at $0.75\pi$. In the nontrivial regime, we find that high-precision sensing is maintained over a broad range of values for the defect parameter $\theta_{02}$, as shown in Fig. \ref{figS2}(a). The Fisher information (FI) consistently exhibits Heisenberg-limited scaling behavior, characterized by a scaling proportional to $t^2$. The FI in the topologically trivial region is consistently lower than that in the nontrivial region, and exhibits a more pronounced decline as the defect $\theta_{02}$ decreases, as shown in Figs.~\ref{figS2}(a–b).

We now fix the defect parameter at $\theta_{02} = -0.55\pi$ to gain a more detailed understanding of the behavior near the topological phase transition point, as shown in Figs.~\ref{figS2}(c–d). Based on this configuration, we further introduce two types of disorder into the QW to assess the robustness of the protocol. Static disorder is introduced by randomly choosing the coin parameters from the interval $[ \theta_i-\pi/20, \theta_i+\pi/20 ]$ at each position, as shown in Figs.~\ref{figS2}(e-f). This type of disorder is spatially dependent but remains constant from step to step.

Dynamic disorder is implemented by randomly varying the coin parameters at each step and uniformly across all positions, within the same interval, as shown in Figs.~\ref{figS2}(g-h). The results show that under both types of disorder, the disorder-averaged FI in the topologically nontrivial region remains consistent with the original distribution along the Heisenberg limit, which demonstrates the robustness. In contrast, within the topologically trivial region, the FI exhibits a more pronounced numerical decline and larger disorder-induced fluctuation compared to the topologically nontrivial case. This demonstrates that the wide-range sensing capability of our protocol is not confined to the variation of $\theta_{02}$ alone, but is also preserved across a broad configuration of $(\theta_1, \theta_2)$ in the topologically nontrivial region, underscoring the generality and robustness of the sensing protocol.

\begin{figure*}
    \centering
    \includegraphics[width=0.75\textwidth]{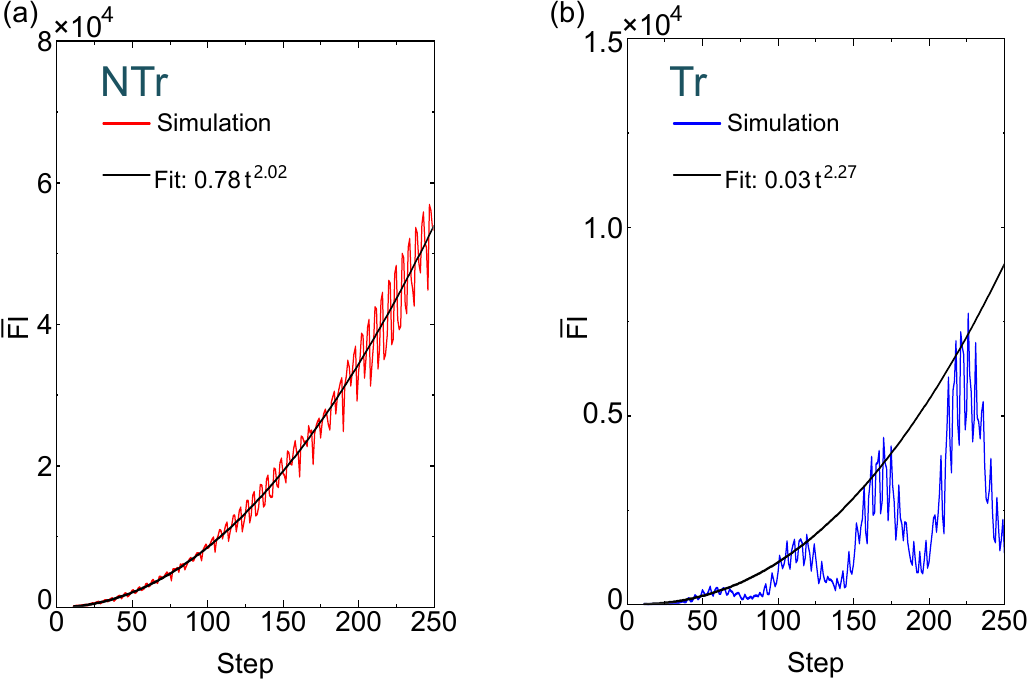}
    \caption{Five-time averaged Fisher information ($\overline{\text{FI}}$) for (a) the topologically nontrivial (NTr) case and (b) the topologically trivial (Tr) case. The parameters are fixed to $\theta_2=0.75\pi$, with $\theta_1=0.90\pi$ in (a) and $\theta_1=0.05\pi$ in (b).}
    \label{figS4}
\end{figure*}

\subsection{Fisher Information for different measurement schemes}

In the previous analysis of the main text, we demonstrate that our protocol achieves high precision and strong robustness in the topologically nontrivial phase, by measuring the FI of finding the walker at the defect site $x = 0$. This local measurement offers the most experimentally feasible route, as it requires monitoring only a single site during the evolution. Since the measurement is confined to the defect site, a portion of the FI distributed the other positions in the system is not captured.

In the topologically nontrivial phase, two topologically protected localized states emerge around the defect, leading to a pronounced concentration of FI at that position. In contrast, in the topologically trivial phase, the absence of such protection allows the FI to spread across the system, thereby diminishing the accessible information.

For comparison, we extend our study to other measurement schemes. As shown in Figs.~\ref{figS3}(a-b), we present the global Fisher information (GFI) obtained from position measurements, which is defined as \begin{equation}\label{GFI}
  \text{GFI}(t)=\sum_i\frac{1}{P_i(t)}\left(\frac{\partial P_i(t)}{\partial\theta_{02}}\right)^2,
\end{equation}
where $P_i(t)$ denotes the probability of finding the walker at the site $x_i$ after $t$ steps. Compared with the results presented in the main text for the same topologically nontrivial and trivial cases, the drop in GFI is less pronounced, while both the quantitative results and the scaling behavior remain unchanged.

In Figs.~\ref{figS3}(c-d),
we present the quantum Fisher information (QFI) obtained from measurements on both coin and position spaces, which is defined as
\begin{equation}\label{QFI}
  \text{QFI}(t)=4\left(\langle\partial_{\theta_{02}}\psi(t)|\partial_{\theta_{02}}\psi(t)\rangle-|\langle\partial_{\theta_{02}}\psi(t)|\psi(t)\rangle|^{2}\right).
\end{equation}
Here $\ket{\psi(t)}$ denotes the quantum state after $t$ steps of evolution from the initial state $|-1,\downarrow\rangle$. Since the QFI incorporates the full information of the evolved states, it exhibits a smooth curve without oscillations. Nevertheless, in the topologically trivial case, the QFI values remain significantly smaller than those in the topologically nontrivial case.

In both cases, the magnitude and scaling show no substantial improvement. Therefore, the FI obtained at the defect site captures the dominant contribution and is sufficient to demonstrate both the key features and the Heisenberg-limit scaling, allowing efficient parameter estimation with minimal measurement resources.

\subsection{Suppressing the Oscillatory Behavior of Fisher Information via multiple-time sampling}

To mitigate the oscillations of FI that arise in the single-time scenario, as shown in the main text, we employ a multiple-time sampling strategy. This approach exploits the additivity of FI. Specifically, we select $K$ distinct evolution times $\{t_1,t_2,\ldots,t_K\}$ and the time points are considered equally spaced with five steps difference between two points, i.e., $t_{i+1}-t_i=5$. We define the average FI as
\begin{equation}\label{avgFI}
        \overline{\text{FI}}(t_{\mathrm{avg}})=\frac{1}{K}\sum_i^K\text{FI}(t_i),
      \end{equation}
where the average evolution time $t_{\mathrm{avg}}=\sum_i^Kt_i/K$, and $\text{FI}(t_i)$ denotes the FI at the defect site $x = 0$ after $t_i$ steps of the QW.

We present the case of $K=5$ as shown in Fig.~\ref{figS4}. In the topologically nontrivial regime, the drop of FI is effectively suppressed by incorporating multiple time points, thereby \textit{improving the estimation precision of the sensor over the entire evolution}. By contrast, the average FI in the topologically trivial regime still exhibits oscillations due to the original fluctuating behavior, combining several time points does not lead to a noticeable smoothing or improvement compared with the single-time case. 

\begin{figure*}
    \centering
    \includegraphics[width=0.75\textwidth]{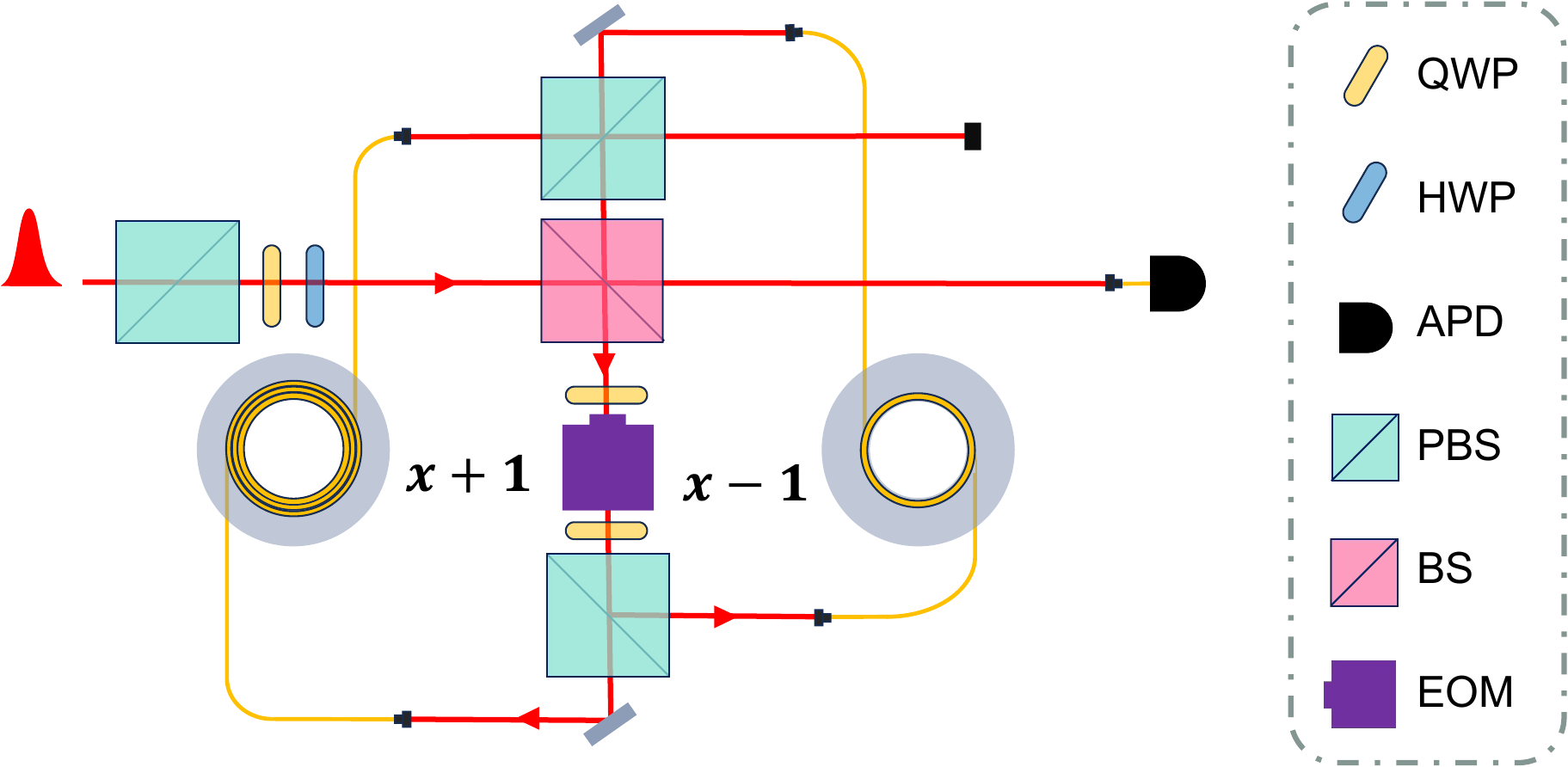}
    \caption{Experimental setup for realizing the quantum walk with a defect.
    State preparation is achieved by subjecting the single photons through a polarizing beam splitter (PBS), a quarter-wave plate (QWP), and a half-wave plate (HWP), after which they are coupled into and out of the interferometric network via a beam splitter (BS). The coin operators are implemented using quarter-wave plates (QWPs) and an electro-optic modulator (EOM).  The shift operator is implemented by directing photons through a PBS into two optical fiber paths of unequal lengths, with the shorter assigned to $x-1$ and the longer to $x+1$. In this configuration, the spatial mode of the walker is encoded via different time delays. The out-coupled photons are detected by avalanche photodiodes (APDs). }
    \label{figS5}
\end{figure*}

\subsection{Experimental proposal}

Based on the topological sensing protocol discussed above, we propose a potential experimental implementation employing a time-multiplexed scheme. As illustrated in Fig.~\ref{figS5}, to achieve the dynamics of this QW with a defect, we construct a fiber-loop interferometric network in which the position of the walker is mapped onto the time domain, while the internal coin state is encoded by the polarization state of the photon. The photons are initially prepared in the specific polarized mode by using a combination of a polarizing beam displacer (PBS), a half-wave plate (HWP), and a quarter-wave plate (QWP). A single step of the QW with the unitary operation \(U=T_{\downarrow}\tilde{R}_2 T_{\uparrow}\tilde{R}_1\), is realized when the photon completes two round trips through the fiber loop. Here the coin operator is realized by the combination of two HWPs with a electro-optical modulator (EOM). By taking advantage of the fast response of the EOM, we can manipulate the polarization state of photons at specific times. This enables the application of coin operators with different parameters at designated position, thereby introducing a defect into the system.

The output of the interferometric network is monitored using a avalanche photodiodes (APDs) to record the time and number of the outgoing photons. This enables the extraction of information not only about the number of steps, but also about the spatial and coin-state distributions of the walker, thereby allowing us to obtain the occupation probability and perform the Bayesian estimation.

\end{widetext}

\end{document}